\documentclass[3p,times,procedia]{elsarticle}
\usepackage{nupha_ecrc}

\usepackage{hyperref}
\usepackage{lineno}

\usepackage{esvect}

\volume{00}

\firstpage{1}

\journalname{Nuclear Physics A}

\runauth{}

\jid{nupha}

\jnltitlelogo{Nuclear Physics A}

\usepackage{amssymb}

\usepackage[figuresright]{rotating}

\begin{document}

\begin{frontmatter}



\dochead{}

\title{Measurements of heavy-flavour nuclear modification factor and elliptic flow in Pb--Pb collisions \\ at $\sqrt{s_{\rm{NN}}}$ = 2.76 TeV with ALICE}


\author{Andrea Dubla, for the ALICE collaboration}

\address{Utrecht University, Princeton Plain 5, Utrecht, The Nederlands }

\begin{abstract}
Heavy quarks, i.e. charm and beauty, are sensitive probes of the medium produced in high-energy heavy-ion collisions. They are produced in the early stage of the collisions and are expected to experience the whole collision evolution interacting with the medium constituents via both elastic and inelastic processes.
The nuclear modification factor ($R_{\rm AA}$) and the elliptic flow ($v_{2}$) are two of the main experimental observables that allow us to investigate the interaction strength of heavy quarks with the medium.
The ALICE collaboration measured the production and elliptic flow of open heavy-flavour hadrons via their hadronic and semi-leptonic decays to electrons at mid-rapidity and to muons at forward rapidity in Pb--Pb collisions.
Recent results will be discussed, and model calculations including the interaction of heavy quarks with the hot, dense, and deconfined medium will be confronted with the data.

\end{abstract}


\end{frontmatter}









\section{Introduction}

The main goal of the ALICE \cite{ALICE2} experiment is to study strongly-interacting matter at high energy density and temperature reached in ultra-relativistic heavy-ion collisions at the Large Hadron Collider (LHC).
In such collisions a deconfined state of quarks and gluons, the Quark-Gluon Plasma (QGP), is expected to be formed.
Due to their large masses, heavy quarks, i.e. charm ($c$) and beauty ($b$) quarks, are produced at the initial stage of the collision, almost exclusively in hard partonic scattering processes. Therefore, they experience the full evolution of the system propagating through the hot and dense medium and loosing energy via radiative~\cite{Radiativea} and collisional~\cite{Colla} scattering processes. Heavy-flavour hadrons and their decay products are thus effective probes to study the properties of the medium created in heavy-ion collisions. 

The nuclear modification factor $R_\mathrm{AA}(p_\mathrm{T}) = \frac{1}{<T_\mathrm{AA}>} \frac{dN_\mathrm{AA}/d\it{p}_{T}}{d\sigma_\mathrm{pp}/{dp_\mathrm{T}}}$ of heavy-flavour hadrons (and their decay leptons) is well established as a sensitive observable to study the interaction strength of hard partons with the medium. Further insight into the medium properties is provided by the measurement of the anisotropy in the azimuthal distribution of particle momenta, that is characterized by the second Fourier coefficient \mbox{$v_{2} = < \cos[2(\varphi - \psi_{2})] >$}. 

Charm and beauty production was measured with ALICE in Pb--Pb collisions at $\sqrt{s_\mathrm{NN}}$  = 2.76 TeV using electrons and muons from semi-leptonic decays of heavy-flavour hadrons and fully reconstructed D-meson hadronic decays. 
D mesons were reconstructed at mid-rapidity ($|y|$ $<$ 0.5) via their hadronic decay channels: ${\rm D}^{0} \rightarrow K^{-}\pi^{+}$, ${\rm D}^{+}\rightarrow K^{-}\pi^{+}\pi^{+}$, ${\rm D}^{*+} \rightarrow {\rm D}^{0}\pi^{+}$ and ${\rm D}^{+}_{s} \rightarrow \phi \pi^{+} \rightarrow K^{-}K^{+}\pi^{+}$ and their charge conjugates. 
The electron identification in the mid-rapidity region \mbox{ ($|y|$ $<$ 0.8)} was based on the d{\it E}/d{\it x} in the TPC. In the low $p_\mathrm{T}$ intervals ($p_\mathrm{T}$ $<$ 3 GeV/{\it c}), where the $K^{\pm}$, proton and deuteron Bethe-Bloch curves cross that of the electron, the measured time-of-flight in TOF and the energy loss in the ITS were employed in addition. At higher $p_\mathrm{T}$, the ratio of the energy deposited in the ElectroMagnetic Calorimeter (EMCal) and the momentum measured with the TPC and ITS, which is close to unity for $\mathrm{e}^{\pm}$, was used to further reject hadrons. Muon tracks were reconstructed in the Forward Muon Spectrometer \mbox{(-4 $<$ y $<$ -2.5)}.

\section{Heavy-flavour production and azimuthal anisotropy}
\begin{figure}[htb]
 \begin{minipage}[b]{0.4\linewidth}
\centering
\includegraphics[height=2.25in]{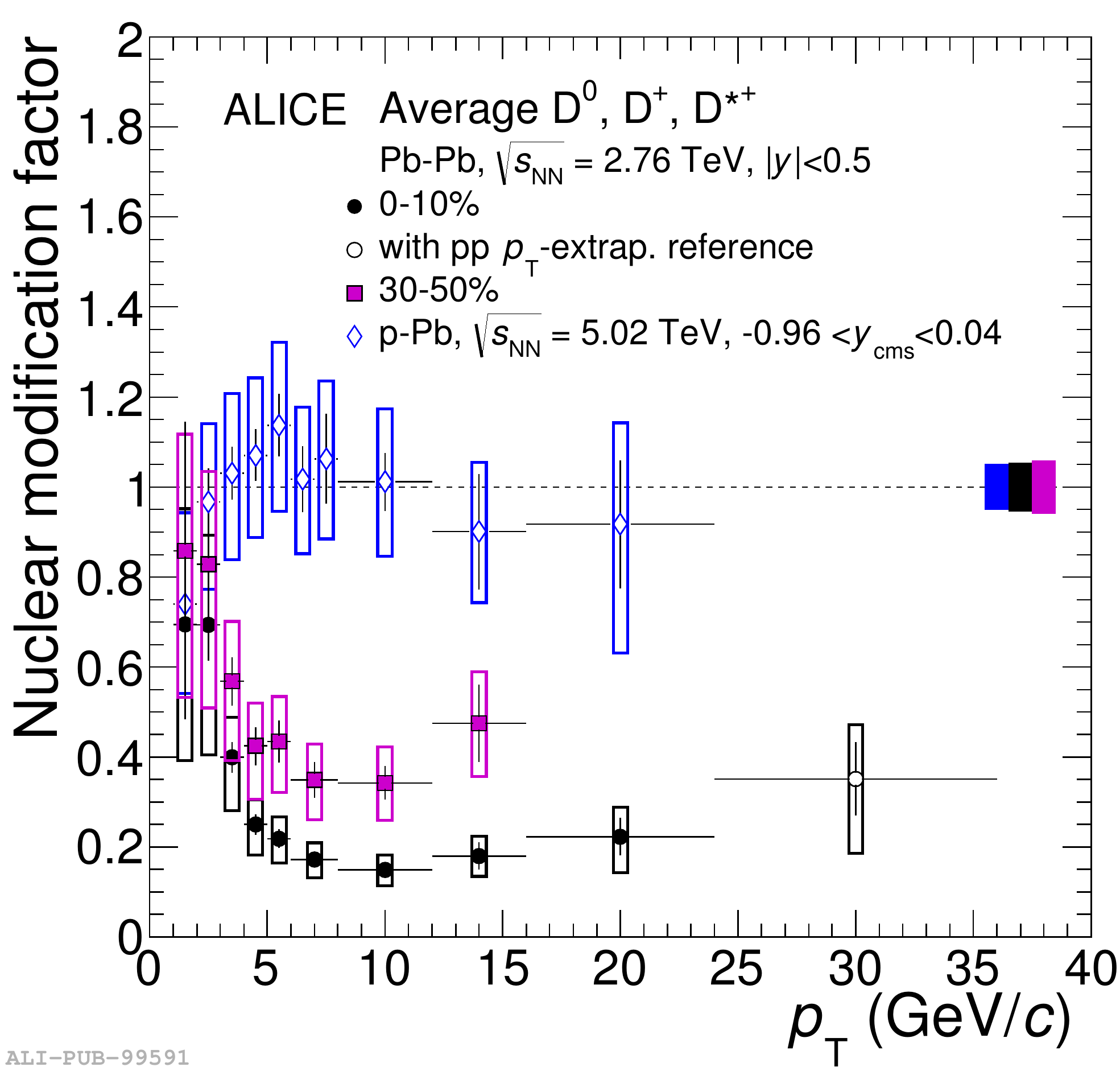}
\end{minipage}
\hspace{2cm}
\begin{minipage}[b]{0.4\linewidth}
\centering
\includegraphics[height=2.25in]{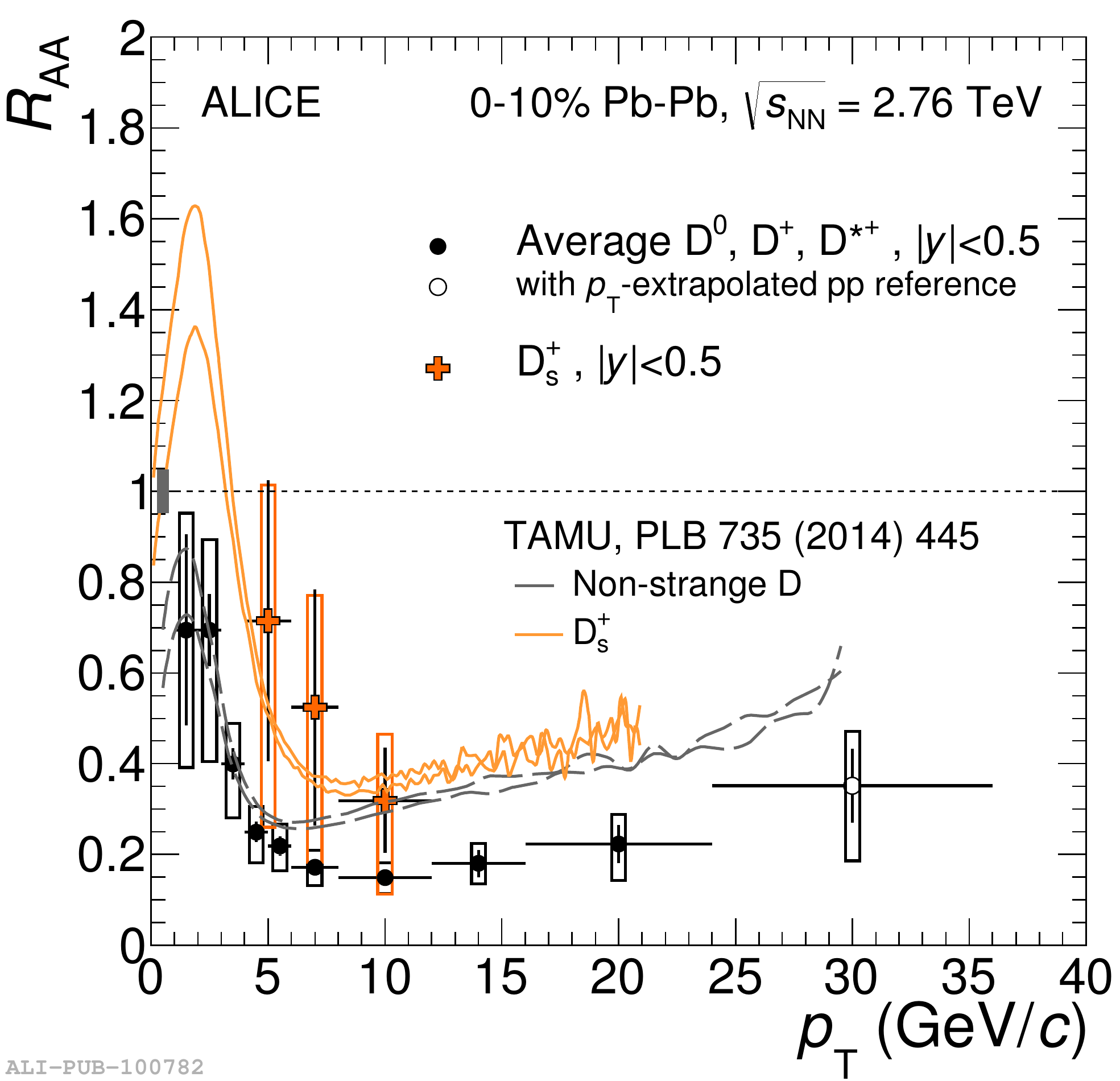}
\end{minipage}
\caption{ Left:  prompt D-meson $R_\mathrm{AA}$ as a  function of $p_\mathrm{T}$ in Pb--Pb collisions at $\sqrt{s_\mathrm{NN}}$  = 2.76 TeV in the 0-10\% and 30-50\% centrality classes~\cite{DMesonRPBPB} and in p--Pb collisions at $\sqrt{s_\mathrm{NN}}$ = 5.02 TeV~\cite{DMesonRpPB}. Right: $R_\mathrm{AA}$ of prompt ${\rm D}^{+}_{s}$ and non-strange D mesons in the 0-10\% centrality class compared to predictions of the TAMU model \cite{TAMU}.}
\label{fig:figure1}
\end{figure}

The ALICE collaboration measured the $R_\mathrm{AA}$ of open heavy-flavour hadrons via their hadronic and semi-leptonic decays in Pb--Pb collisions at $\sqrt{s_\mathrm{NN}}$ = 2.76 TeV~\cite{MuonRAA, DMesonRPBPB}.
The left panel of Fig.~\ref{fig:figure1} shows the nuclear modification factor of prompt D mesons (average of ${\rm D}^{0}$, ${\rm D}^{+}$ and ${\rm D}^{*+}$) as a function of $p_\mathrm{T}$ in central (0-10\%) and semi-central (30-50\%) Pb--Pb and in p--Pb collisions~\cite{DMesonRPBPB, DMesonRpPB}. A stronger suppression is observed in more central with respect to semi-central Pb--Pb collisions in the measured $p_\mathrm{T}$ range.
An $R_\mathrm{pPb}(p_\mathrm{T})$ consistent with unity was measured for \mbox{$p_\mathrm{T}$ $>$ 3 GeV/$c$}, confirming that the suppression observed in central \mbox{Pb--Pb} collisions is predominantly induced by final-state effects due to charm quark energy loss in the medium.

\begin{figure}[htb]
 \begin{minipage}[b]{0.4\linewidth}
\centering
\includegraphics[height=2.25in]{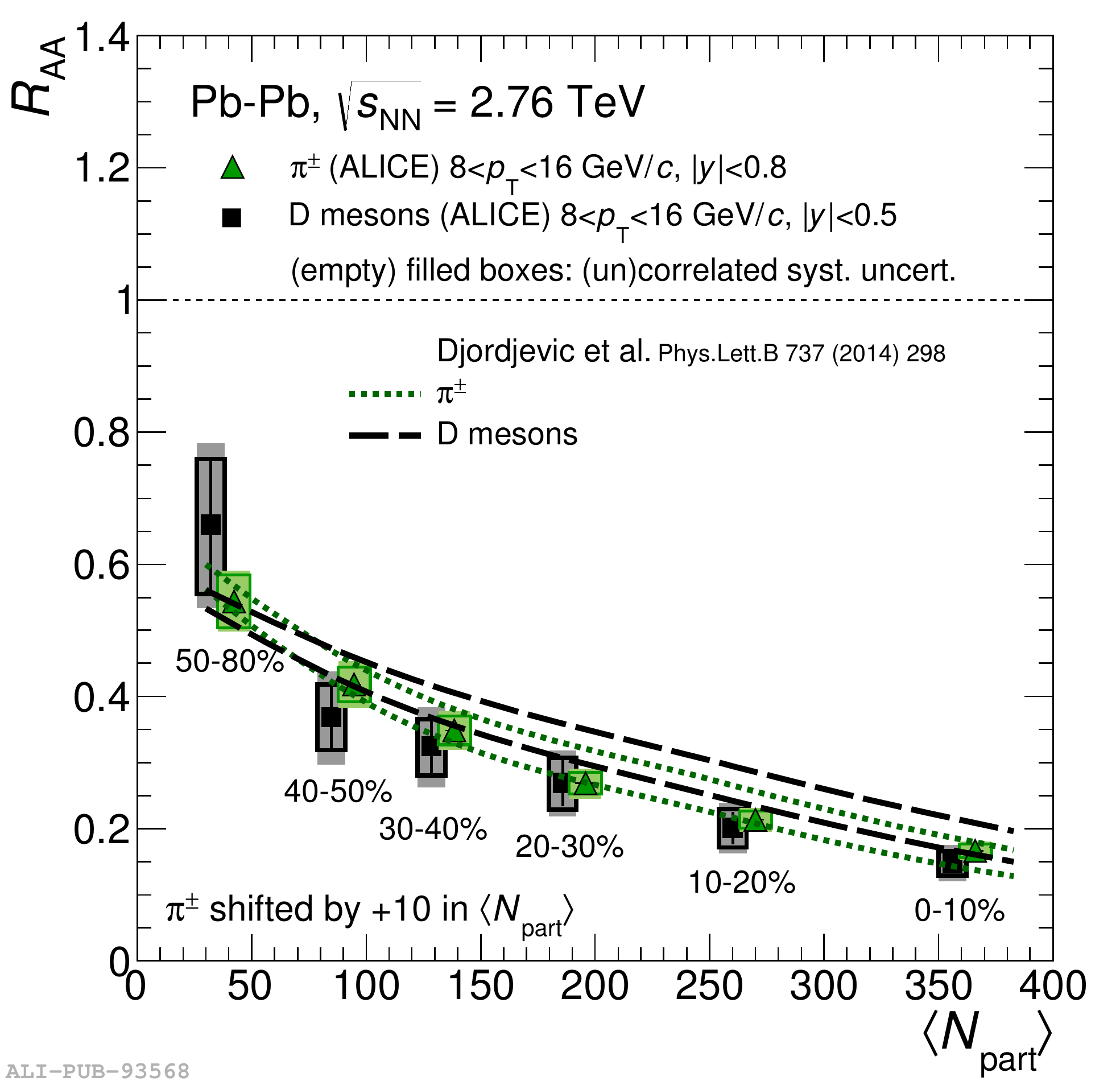}
\end{minipage}
\hspace{2cm}
\begin{minipage}[b]{0.4\linewidth}
\centering
\includegraphics[height=2.25in]{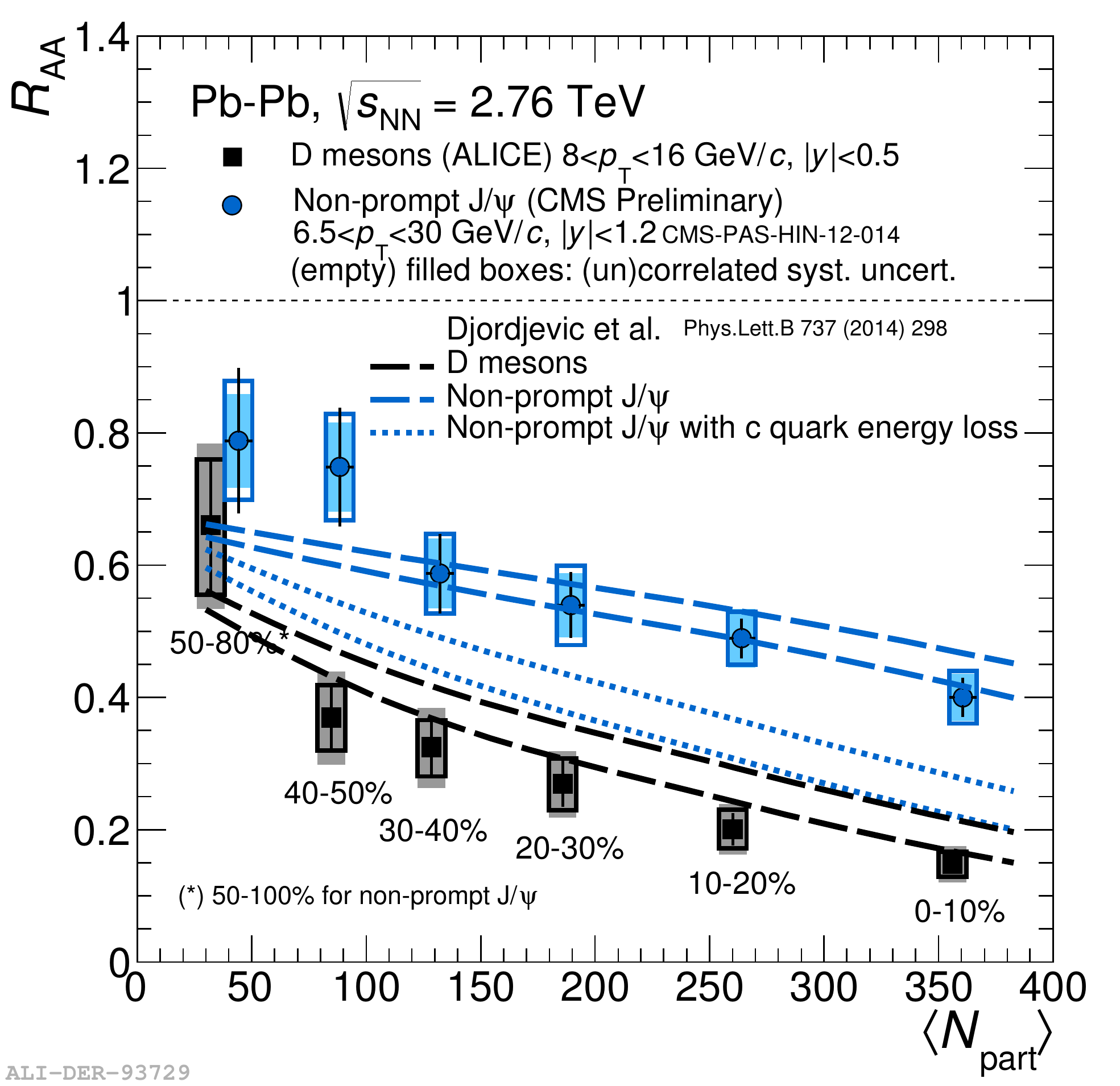}
\end{minipage}
\caption{Comparison of the $R_\mathrm{AA}$ measurements as a function of centrality with the model calculations~\cite{Djordjevic} including radiative and collisional energy loss. Left: D mesons and charged pions in 8 $<$ $p_\mathrm{T}$ $<$ 16 GeV/{\it c}~\cite{DmesonRAACent, RAAPion}. Right: D mesons in 8 $<$ $p_\mathrm{T}$ $<$ 16 GeV/{\it c} and non-prompt J/$\Psi$ mesons in 6.5 $<$ $p_\mathrm{T}$  $<$ 30 GeV/$c$ \cite{RAACMS}.}
\label{fig:figure2}
\end{figure}

In addition the ${\rm D}^{+}_{s}$ meson is sensitive to strangeness production and to the hadronization mechanism of charm quarks. 
Due to the strangeness enhancement in QGP, the relative yield of ${\rm D}^{+}_{s}$ mesons with respect to non-strange charmed mesons at low $p_\mathrm{T}$ is predicted to be enhanced in nucleus-nucleus collisions as compared to pp interactions \cite{hadmodel1}.
In the right panel of Fig.~\ref{fig:figure1}, the $p_\mathrm{T}$-differential $R_{\rm AA}({\rm D}_{s}^{+})$ is compared with the $R_{\rm AA}({\rm D})$ of non-strange D mesons. Strong ${\rm D}^{+}_{s}$ suppression in central collisions (similar to non-strange D meson) was measured for 8 $<$ $p_\mathrm{T}$ $<$ 12 GeV/$c$ with a hint of less suppression for $p_\mathrm{T}$ $<$ 8 GeV/$c$~\cite{DSMesonRPBPB}.
The TAMU \cite{TAMU} model describes the measured ${\rm D}^{+}_{s}$ nuclear modification factor within uncertainties, and it also provides a reasonable description of non-strange D-meson $R_\mathrm{AA}$ at low $p_\mathrm{T}$.
Due to the QCD nature of parton energy loss, quarks are predicted to lose less energy than gluons (that have a larger colour coupling factor). In addition, the dead-cone effect \cite{DeadCone} is expected to reduce the energy loss of massive quarks with respect to light quarks.
Therefore, if we consider only energy-loss effect, a hierarchy in the $R_\mathrm{AA}$ is expected to be observed when comparing the mostly gluon-originated light-flavour hadrons (e.g. pions) to D and to B mesons \cite{DMesonRPBPB, BMe}: \mbox{$R_{\rm AA}({ \pi})$ $<$ $R_{\rm AA}({\rm D})$ $<$ $R_{\rm AA}({\rm B})$}. The measurement and comparison of these different medium probes provides a unique test of the colour-charge and mass dependence of parton energy loss.
In the left panel of Fig. \ref{fig:figure2} the nuclear modification factor of prompt D mesons in the transverse momentum region 8 $<$ $p_\mathrm{T}$ $<$ 16 GeV/{\it c} is shown as a function of centrality in comparison with the $R_{\rm AA}({ \pi})$. Less suppression is observed moving from central to semi-central collisions, since the medium formed in peripheral collisions should be less dense with respect to the one formed in a central collision. The D-meson $R_{\rm AA}({\rm D})$ is compatible with that of charged pions and charged particles within uncertainties \cite{DMesonRPBPB, DmesonRAACent, RAACH, RAAPion}. The consistency between the two measurements is also described by a model including mass-dependent energy loss, different shape of the parton $p_\mathrm{T}$ spectra and different parton fragmentation functions~\cite{Djordjevic}.

The comparison of the centrality dependence of the $R_\mathrm{AA}$ of D mesons and of J/$\Psi$ from B-hadron decays (measured by the CMS collaboration \cite{RAACMS}) is displayed in the right panel of Fig.~\ref{fig:figure2}. It shows an indication for a stronger suppression for charm than for beauty at high $p_\mathrm{T}$ in central Pb--Pb collisions consistent with the expectation of $R_{\rm AA}({\rm D})$ $<$ $R_{\rm AA}({\rm B})$. 
The two measurements are described by the predictions based on a pQCD model including mass-dependent radiative and collisional energy loss \cite{Djordjevic}. In this model the difference in the $R_\mathrm{AA}$ of charm and beauty mesons is mainly due to the mass dependence of quark energy loss, as demonstrated by the curve in which the non-prompt J/$\Psi$ $R_\mathrm{AA}$ is calculated assuming that $b$ quarks suffer the same energy loss as $c$ quarks. 

\begin{figure}[htb]
\centering
\includegraphics[height=2.3in]{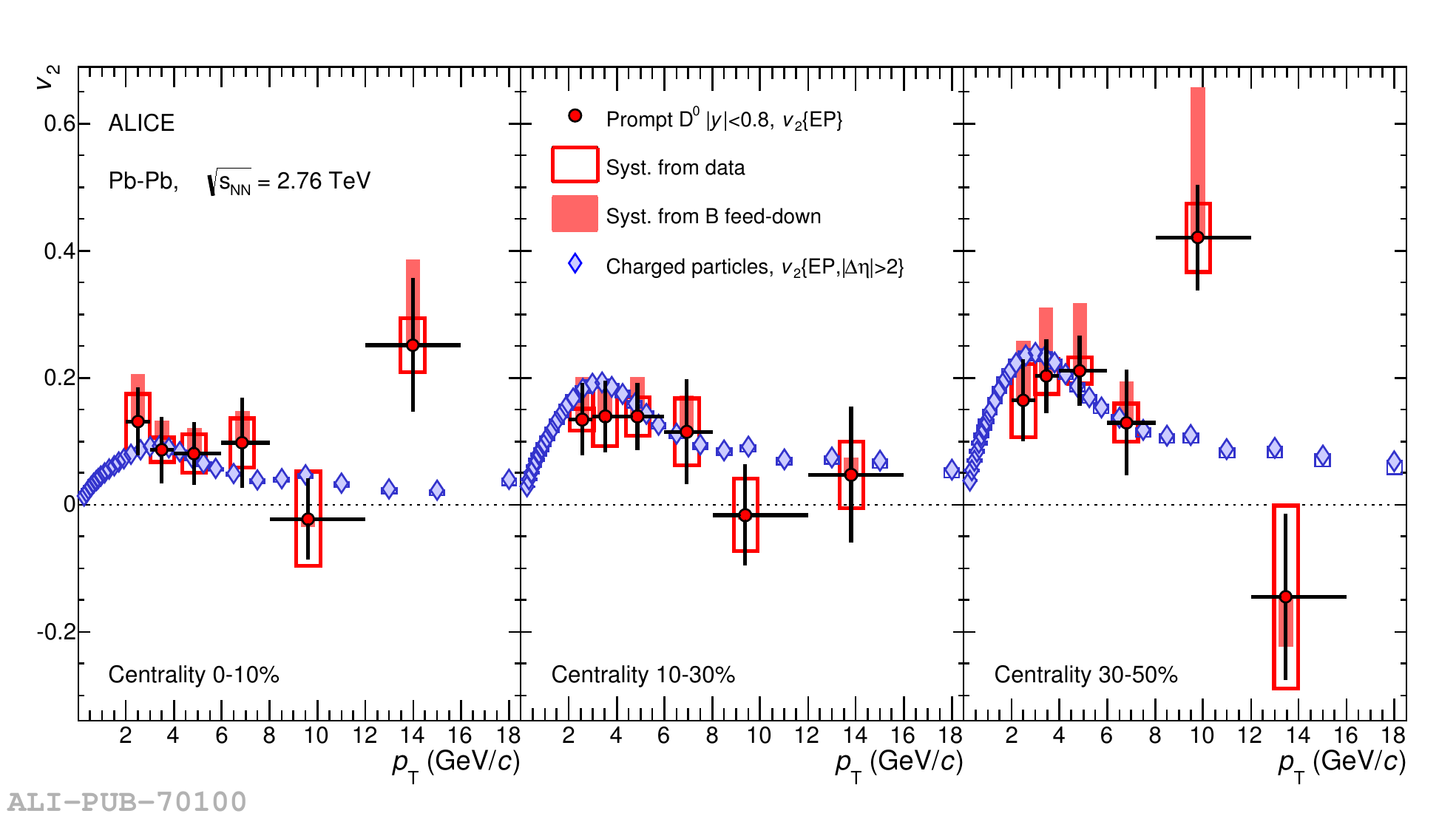}
\caption{$p_\mathrm{T}$-differential $v_2$ of ${\rm D}^{0}$ mesons and charged particles in Pb--Pb collisions at $\sqrt{s_\mathrm{NN}}$  = 2.76 TeV for three centrality intervals~\cite{DMESONV2Long}. }
\label{fig:figureDv2}
\end{figure}

\begin{figure}[htb!]
\begin{minipage}[b]{0.33\linewidth}
\centering
\includegraphics[height=2.1in]{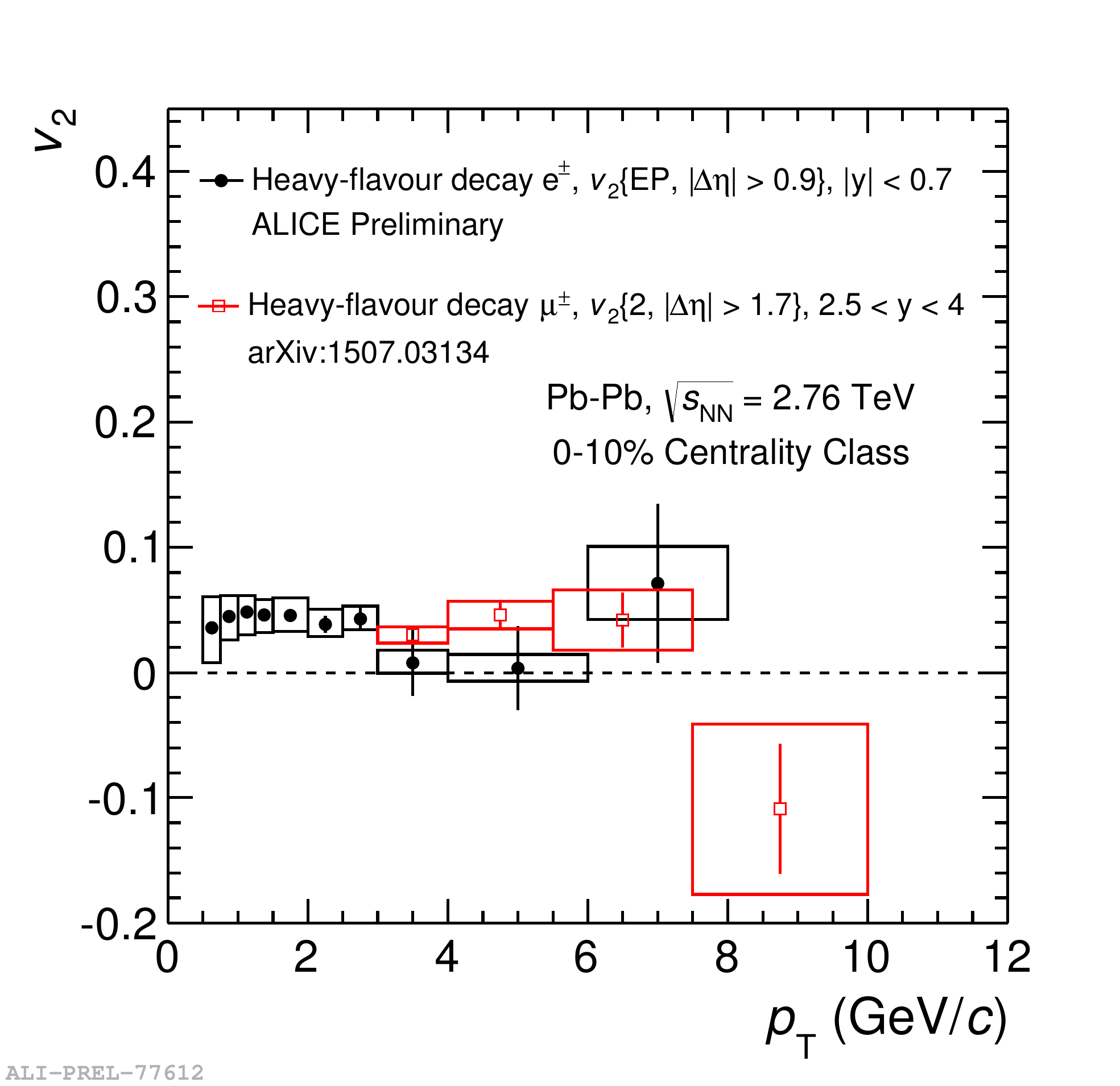}
\end{minipage}
 \begin{minipage}[b]{0.33\linewidth}
\centering
\includegraphics[height=2.1in]{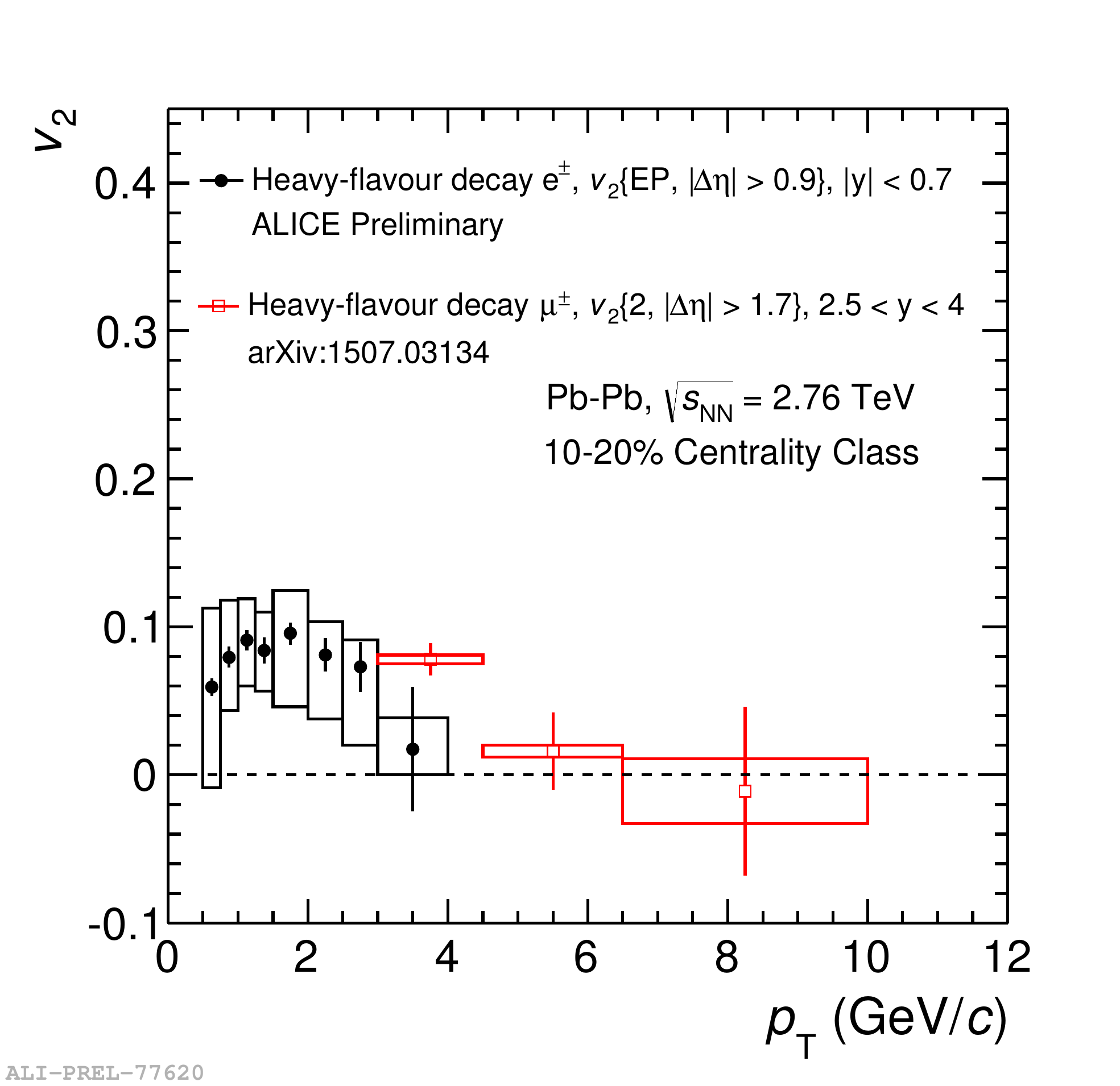}
\end{minipage}
 \begin{minipage}[b]{0.0\linewidth}
\centering
\includegraphics[height=2.1in]{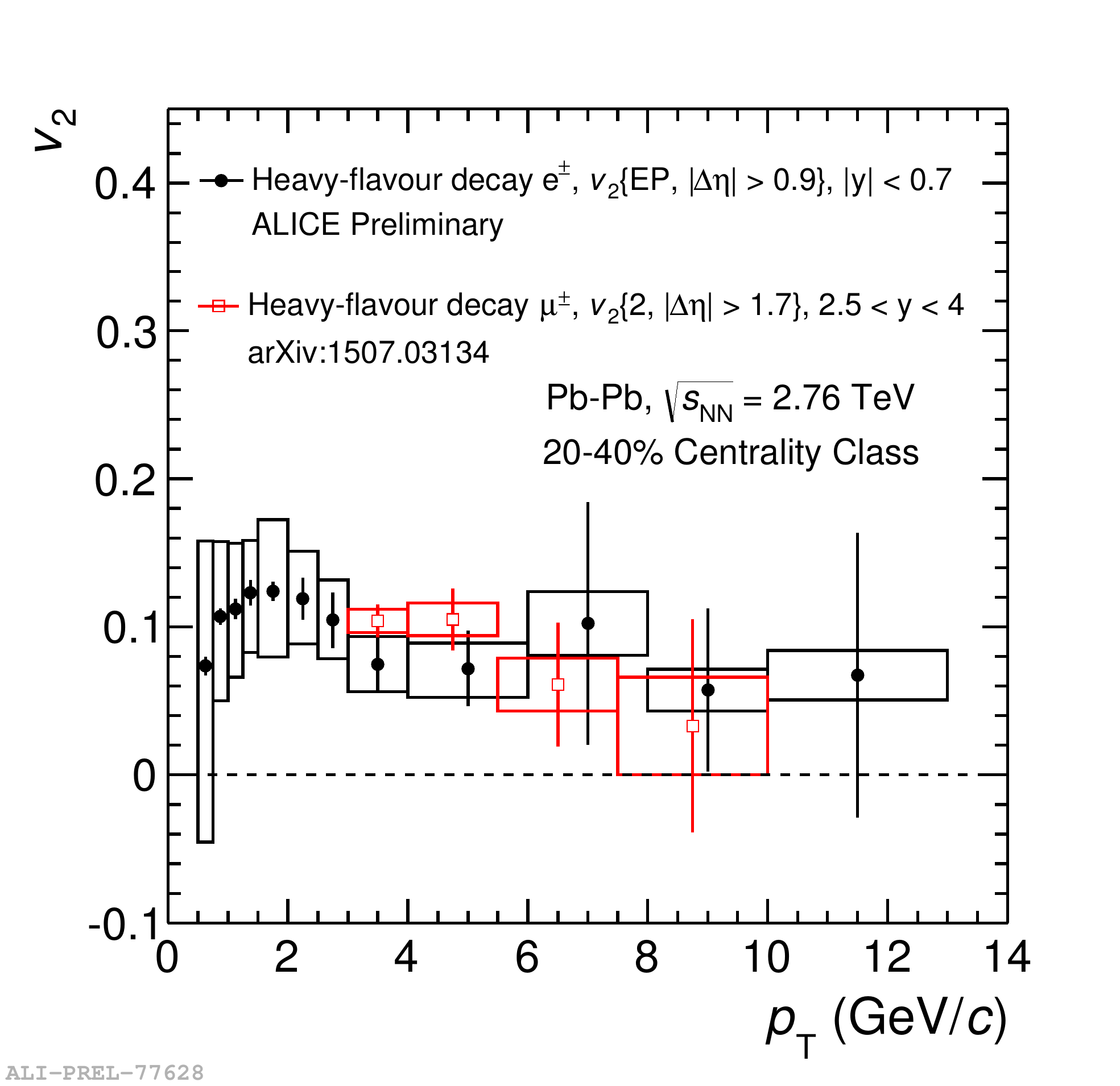}
\end{minipage}
\caption{$p_\mathrm{T}$-differential elliptic flow of heavy-flavour hadron decay electrons at mid-rapidity and muons at forward rapidity measured with the event-plane method and with two-particle Q-cumulant, respectively, in Pb--Pb collisions at $\sqrt{s_\mathrm{NN}}$ = 2.76 TeV for three centrality intervals.}
\label{fig:figure8}
\end{figure}

The ALICE collaboration also measured the elliptic flow $v_{2}$ of open heavy-flavour hadrons via their hadronic and semi-leptonic decays in Pb--Pb collisions at $\sqrt{s_\mathrm{NN}}$ = 2.76 TeV~\cite{DMESONV2,Muon, HFEv2, DMESONV2Long}.
The measured averaged $v_{2}$  of prompt ${\rm D}^{0}$, ${\rm D}^{+}$ and ${\rm D}^{*+}$  indicates a positive $v_{2}$ in semi-central (30-50\%) Pb--Pb collisions with a significance of 5.7 $\sigma$ for 2 $<$ $p_\mathrm{T}$ $<$ 6 GeV/{\it c} \cite{DMESONV2}.
The anisotropy of prompt  ${\rm D}^{0}$ mesons was measured in the three centrality classes 0-10\%, 10-30\% and 30-50\%, as reported in Fig.~\ref{fig:figureDv2}. The results show a hint of increasing $v_{2}$ from central to semi-peripheral collisions and are comparable in magnitude to that of inclusive charged particles, suggesting that charm quarks participate in the collective flow of the expanding medium~\cite{DMESONV2, DMESONV2Long}.
The elliptic flow of heavy-flavour hadron decay electrons at mid-rapidity~\cite{HFEv2} and muons at forward rapidity~\cite{Muon} was measured in the three centrality classes 0-10\%, 10-20\% and 20-40\%. The measurements are comparable in magnitude as shown in Fig.~\ref{fig:figure8}. The magnitude of $v_{2}$ increases from central to semi-central collisions. In semi-central (20-40\%) Pb--Pb collisions a positive $v_{2}$ is observed with a significance of 3$\sigma$ for 2 $<$ $p_\mathrm{T}$ $<$ 3 GeV/{\it c} for the electrons and for 3 $<$ $p_\mathrm{T}$ $<$ 5 GeV/{\it c} for the muons. These results confirm the significant interaction of heavy quarks, mainly charm, with the medium. 


The results obtained with ALICE using the data from the LHC Run-1 (2010-2013) indicate a strong modification of heavy-flavour production in central Pb--Pb collisions, with a clear difference in the D and B-meson suppression at high $p_\mathrm{T}$ as expected by the quark-mass dependence of the energy loss. Elliptic flow measurements obtained with ALICE in semi-central Pb--Pb collisions at $\sqrt{s_\mathrm{NN}}$ = 2.76 TeV suggest a collective motion and a possible thermalization of low-$p_\mathrm{T}$ heavy quarks in the medium, mainly charm. Several models including interactions of the $c$ and $b$ quarks with a hot, dense and deconfined medium can qualitatively describe the features observed in the data.
A simultaneous description of the  $R_\mathrm{AA}$ and $v_{2}$ starts to provide constraints to the models themselves.

\bibliographystyle{elsarticle-num}
\bibliography{<your-bib-database>}



\end{document}